\documentclass{llncs}
\usepackage{graphicx}
\usepackage{amsmath}
\newcommand{\eps}{\epsilon}
\newcommand{\XOR}{\oplus}
\newcommand{\de}{\partial}
\newcommand{\A}{\mathcal{A}}
\newcommand{\B}{\mathcal{B}}

\begin{document}

\pagestyle{empty}

\mainmatter

\title{Opinion formation and phase transitions in a probabilistic cellular
 	automaton with two absorbing states}
\author{	
	Franco Bagnoli\inst{1,4} \and  
	Fabio Franci\inst{2,4} \and 
	Ra\'ul Rechtman\inst{3}
}

\institute{
	(Dipartimento di Energetica, 
	Universit\`a di  Firenze, \\ Via S. Marta, 3 
	I-50139 Firenze, Italy.
	\email{bagnoli@dma.unifi.it} \\
	\and
	Dipartimento di  Sistemi e Informatica, 
	Universit\`a di  Firenze, \\ Via S. Marta, 3 
	I-50139 Firenze, Italy.
	\email{bagnoli@dma.unifi.it} 
	\and
  Centro de Investigac\'\i{}on en Energ\'\i{}a, 
 UNAM, \\62580 Temixco, Morelos, Mexico. 
 \email{rrs@teotleco.cie.unam.mx}
 \and
 INFM, Sezione di Firenze.
}

\maketitle
\setcounter{footnote}{0}

\begin{abstract}
We discuss the process of opinion formation in a completely homogeneous, democratic 
population using a class of probabilistic cellular automata models with two absorbing
states. Each individual can  have one of two opinions that can change
according to that of his  neighbors. It is dominated by an
overwhelming majority and can disagree against a marginal one. We
present the phase diagram in the mean field approximation  and from
numerical experiments for the simplest nontrivial case. For
arbitrarily large neighborhoods we  discuss the mean field results
for a non-conformist society, where each individual adheres to the
overwhelming majority of its neighbors and choses an opposite
opinion  in other cases. Mean field results and preliminary lattice simulations with long-range connections among individuals
show the presence of coherent temporal oscillations of the population.
\end{abstract}

\section{Modeling social pressure and political transitions}

What happens to a society when a large fraction of people switches
from a conformist to a non-conformist attitude?  
Is the transition smooth or
revolutionary?  These important questions, whose answers can make the 
difference between two well-known political points of view, 
is approached using a theoretical
model, in the spirit of Latan\'e's social impact theory~\cite{latane,galam1}.

We assume that one's own inclination towards political choices 
originates from a mixture of a certain degree of 
conformism and non-conformism. Conformists
tend to agree with the local community majority, that is with the average opinion 
in a given neighborhood,  while non-conformists do the opposite. However, an overwhelming majority in the neighborhood
(which includes the subject itself) is always followed. 

We shall study here the case of a homogeneous population, i.e.\ a
homogeneous democratic society.\footnote{The case with strong leader
was studied in Ref~\protect\cite{leader}.} It may be
considered as the annealed version of a real population, which is 
supposedly composed by a
mixture of conformist and non-conformist people who do not change
easily their attitude. 

In Sec.~\ref{sec:model} we introduce a class of 
probabilistic cellular automata characterized 
by the size $2r+1$ of the neighborhood, a majority threshold $q$, 
a coupling constant $J$ and an external field $H$. 

We are interested in the two extreme cases: 
people living on a one-dimensional lattice, interacting only with their nearest neighbors ($r=1$) and 
people interacting with a
mean-field opinion.\footnote{Related models in one and two dimensions have been studied in Refs~\protect\cite{galam2,galam3}.}

In Sec.~\ref{sec:r1} we present the simplest case where each individual interacts with 
his two nearest neighbors ($r=1$), the 
mean field phase diagram and the one found from numerical experiments. 
For this simple case, we find a complex behavior which includes first 
and second order
phase transitions, a chaotic region and the presence of two universality 
classes~\cite{nature}. In Sec.~\ref{sec:meanfield} we discuss the mean
field behavior of the model for arbitrary neighborhoods and 
majority thresholds when the external field is zero and the coupling
constant is negative (non-conformist society). The phase diagram of the model exhibits a large region of coherent temporal oscillations of the whole populations, either chaotic or regular. These oscillations are still present in the lattice version with 
a sufficient large fraction of long-range 
connections among individuals, 
due to the small-world effect~\cite{small}. 

\section{The model}\label{sec:model}

We denote by $x_i^t$ the opinion assumed by individual $i$ at time $t$. We shall limit to two opinions, denoted by $-1$ and
$1$ as usual in spin models.
The system is composed by $L$ individuals arranged on a one-dimensional lattice. All
operations on spatial indices are assumed to be modulo $L$ (periodic
boundary conditions). The time is considered discontinuous
(corresponding, for instance, to election events). The state of the whole system at time
$t$ is given by $\boldsymbol{x}^t=(x_0^t,\dots,x_{L-1}^t)$ with
$x_i^t\in \{-1,1\}$;

The individual opinion is formed according to a local community ``pressure'' and a global influence. 
In order to avoid a tie, we assume that the local community 
is formed by $2r+1$ individuals, counting on equal ground the opinion
of the individual himself at previous time.  
The average opinion of the local community around site $i$ at time $t$ is denoted by $m^t_i = \sum_{j=- r}^{r} x^t_{i+j}$. 

The control parameters are the probabilities $p_s$ of
choosing opinion $1$ at time $t+1$ if this opinion is shared by
$s$ people 
in the local community, 
i.e.\  if the local ``field'' is $m = 2s-2r-1$. 

Let $J$ be a parameter controlling the influence of the local field in the opinion
formation process and  $H$ be the external social
pressure. 
The probability $p_s$ are given by  
\[
	p_s = p_{(m+2r+1)/2}  \propto \exp(H+J m).
\]

One could think to $H$ as the television influence, and $J$ as
educational effects. $H$  pushes towards one opinion or the other, and
people educated towards conformism will have $J>0$, while
non-conformists will have $J<0$. 
In the statistical mechanics lingo, all parameters are rescaled to
include the temperature.

The hypothesis of 
alignment to overwhelming local majority 
is represented by a parameter $q$, indicating the critical size of local majority. If $s < q$ ($m<2q-2r-1$), then $x_i^{t+1}=-1$,
and if $s > 2r+1-q$ ($m>2r+1-2q$), then $x_i^{t+1}=1$.   

In summary, the local transition probabilities of the model are
\begin{equation}\label{p}
p_s = \begin{cases}
 0 & \text{if $s<q$;}\\
 \dfrac{\A\B^s}{1+\A\B^s} & 
 	\text{if $q \le s \le 2r+1-q$;}\\
 1 & \text{if $s>2r+1-q$;}
\end{cases}
\end{equation}
where $\A = \exp[2H+2J(2r-1)]$ and $\B = \exp(4J)$. 

For $q=0$ the model reduces to an Ising spin system. 
For all $q>0$ we have two absorbing homogeneous states,
$\boldsymbol{x}=\boldsymbol{-1}$ ($c=0$) and 
 $\boldsymbol{x}=\boldsymbol{1}$ ($c=1$)
corresponding to infinite coupling (or zero temperature) in the statistical mechanical sense. 
With these assumptions, the model reduces to 
a one-dimensional, 
one-com\-ponent, totalistic cellular automaton with two absorbing
states.

The order parameter is the
fraction $c$ of people sharing opinion $1$.~\footnote{The usual order parameter for magnetic system is the magnetization $M = 2c-1$.} 
It is zero or one  
in the two absorbing states, and assumes
other values in the active phase. The model is symmetric since the  two absorbing states have the same importance.

\section{A simple case} \label{sec:r1}

Let us study the model for the simplest, nontrivial case: one dimensional lattice, $r=1$ and $q=1$. 
We have a probabilistic cellular automaton with three
inputs, and two free control parameters, $p_1$ and $p_2$, while $p_0\equiv 0$ and $p_3\equiv 1$, according to Eq.~\eqref{p}.

One can also invert the relation between $(H,J)$ and $(p_1, p_2)$: 
\[
	H = \dfrac{1}{4} \log\left(\dfrac{p_1p_2}{(1-p_1)(1-p_2)}\right),\qquad
	J = \dfrac{1}{4} \log\left(\dfrac{p_2(1-p_1)}{p_1(1-p_2)}\right).
\]
The diagonal $p_2=p_1$ corresponds to $J=0$ and the diagonal
$p_2=1-p_1$ to $H=0$. 

At the boundaries of probability intervals, we have four deterministic
(elementary) cellular automata, that we denote T3, T23, T13 and
T123, where the digits indicate the values of $s$ for which $p_s=1$~\cite{vichniac}.  

Rule T3 (T123) brings the system into the $c=0$ ($c=1$) absorbing state
except for the special case of the initial configuration homogeneously
composed by the opposite opinion. 

\begin{figure}[t]
\centerline{\includegraphics[width=6cm]{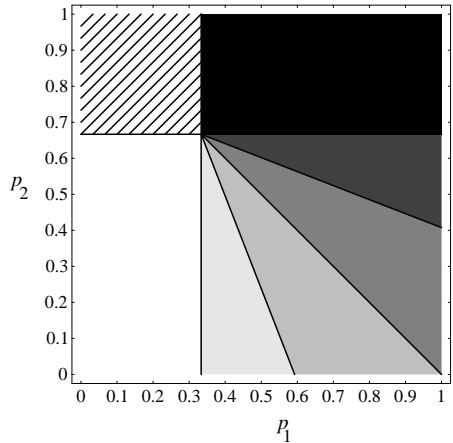}}
\caption{
Mean-field phase diagram for the density 
 $c$ coded as gray levels
 ranging from white ($c=0$) to black ($
 c=1$). The dashed upper-left 
 region denotes the coexistence phase, in which both
 states $c=0$ and $c=1$ are stable, and the final state depends
 on the initial density (first-order transition). 
}
\end{figure}

Rule T23 is a strict majority rule, whose evolution starting from a
random configuration leads to the formation of frozen patches
of zeros and ones in a few time steps. 
A small variation of the
probabilities induces fluctuations in the position of the patches.
Since the patches disappear when the boundaries collide, the system
is equivalent to a model of annihilating random walks, which, in one
dimensions, evolves towards one of the two absorbing states according
to the asymmetries in the probabilities or in the initial configuration. 

Rule T13 on the other hand is a ``chaotic'' one,~\footnote{Called  rule 150 in Ref.~\protect\cite{wolfram}} leading
to irregular pattern for almost all initial conditions (except for the
absorbing states). These patterns
are microscopically very sensitive to perturbations, but very robust
for what concerns global quantities (magnetization). 

The role of frustrations (the difficulty of choosing a stable opinion) is evidentiated by the following considerations.  
Rule T23 corresponds to
a ferromagnetic Ising model at zero temperature, so an individual can simply align with the local majority with no frustrations.
On the contrary, rule T13 is unable to converge towards an absorbing
state (always present), because these states are unstable: a single
individual disagreeing with the global opinion triggers a flip in all
local community to which he/she belongs to. It is possible to quantify
these concepts by defining stability parameters similar to Lyapunov
exponents~\cite{chaos}.


We start by studying the mean-field approximation for the generic case, with
$p_1$ and $p_2$ different from zero or one. 
 
Let $c$ and $c'$ denote the density of
opinion $1$ at times $t$ and $t+1$ respectively. We have 
\[ 
 c' =3p_1c\left(1-c\right)^2+3p_2c^2\left(1-c\right)+c^3.
\]
This map has three fixed points, the state $\boldsymbol{x}=\boldsymbol{-1}$ ($c=0$), the state $\boldsymbol{x}=\boldsymbol{1}$ ($c=1$) 
and a microscopically disordered state ($0\le c \le 1$).
The
model is obviously symmetric under the changes $p_1\to 1-p_2$, $p_2\to 1-p_1$
and $x\to 1-x$, implying a fundamental equivalence of political
opinions in this model. The stability of fixed points marks the different phases, as shown in Fig.~1.

The stability of the state $c=1$ ($c=0$)
corresponds to large social
pressure towards opinion~$1$ ($-1$). The value of $J$
determines if a change in social pressure corresponds to a smooth or
abrupt transition. 

\subsection{Phase transitions and damage spreading in the lattice 
case}

The numerical phase diagram of the
model starting from a random initial state with $c^0=0.5$ 
is shown in Fig.~2. The scenario is qualitatively the same
as predicted by the mean-field analysis. In the upper-left part of the diagram both states $c=0$ and $c=1$ are stable. In this region the final fate of the system depends on the initial configuration.

\begin{figure}[t]
\begin{tabular}{cc}
\includegraphics[width=6cm]{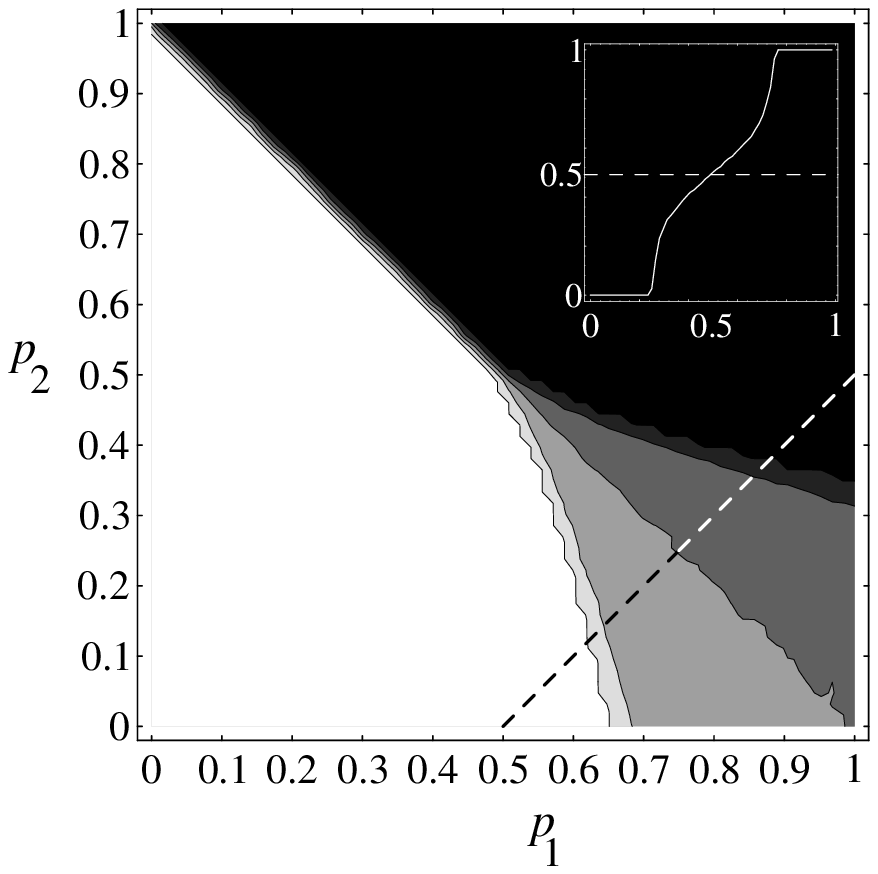} & 
\includegraphics[width=6cm]{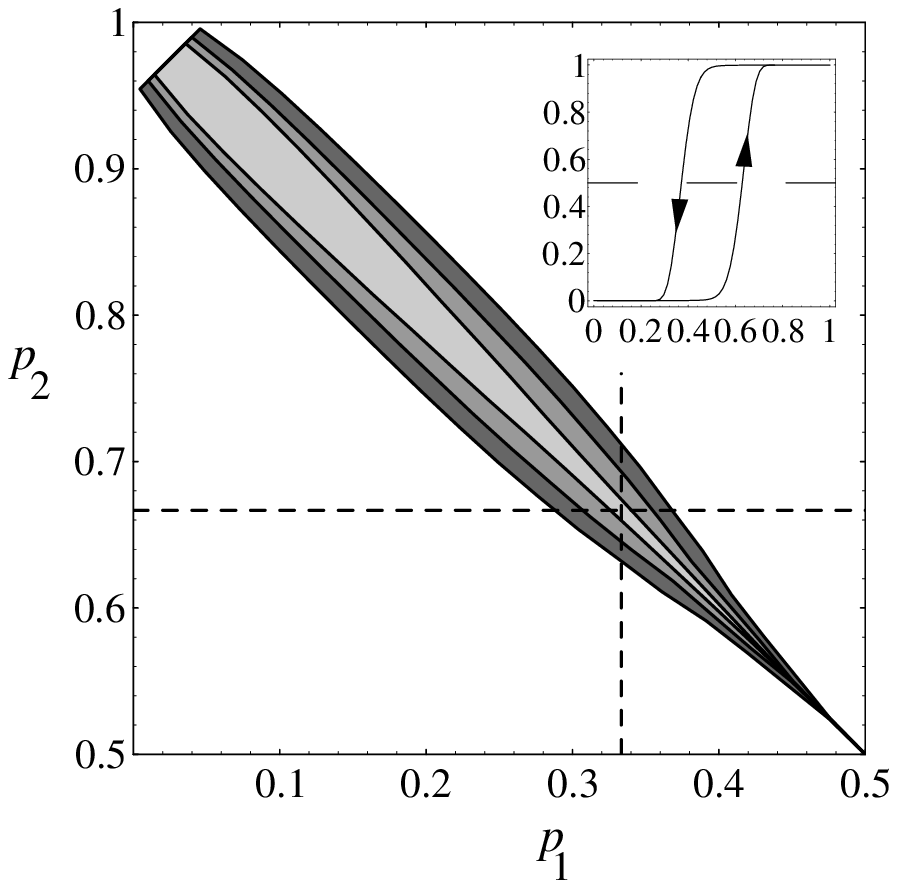}
\end{tabular}
\caption{Numerical phase diagram for the density 
 $c$ (left) and 
hysteresis region for several values of the noise $\eps$
 and relaxation time $T$ (right). Same color code as in Figure~1.}
\end{figure}

Due to the symmetry of the model, the two second-order phase
transition curves meet at a bicritical point $(p_t,1-p_t)$ where
the first-order phase transition line ends.
Crossing the second-order phase boundaries on a line parallel to the
diagonal $p_1=p_2$, the density $c$ exhibits two critical
transitions, as shown in the inset of the right panel of Fig.~2.
Approaching the bicritical point the critical region becomes
smaller, and corrections to scaling increase. Finally, at the
bicritical point, the two transitions coalesce into a single
discontinuous (first-order) one. 

First-order phase transitions are usually associated to a \emph{hysteresis
cycle} due to the coexistence of two stable states. 
To visualize the hysteresis loop (inset of the right panel of Fig.~2) we
modify the model slightly by letting 
$p_0=1-p_3=\eps$ with $\eps \ll
1$. In this way the configurations $\boldsymbol{x}=\boldsymbol{-1}$ and $\boldsymbol{x}=\boldsymbol{1}$ are no 
longer absorbing. This brings the model back into the class of
equilibrium models for which there is no phase transition in one
dimension but metastable states can nevertheless persist for long
times. The width of the hysteresis cycle, shown in the right panel of Fig.~2, depends on the 
value of $\eps$ and the relaxation time $T$.

We study the asymptotic density as $p_1$ and $p_2$ move on a line
with slope 1 inside the dashed region of Fig.~1.
For $p_1$ close to zero, the model has only one stable state, 
close to the state $c=0$. As $p_1$ increases
adiabatically, the new asymptotic density will still assume this
value even when the state $c=1$ become stable. Eventually the first state 
become unstable, and the asymptotic density jumps to the stable
fixed point close to the state $c=1$. Going backwards on the same line, the
asymptotic density will be close to one until that fixed point
disappears and it will jump back to a small value close to zero.

Although not very interesting from the point of view of opinion
formation models, the problem of universality classes in the presence
of absorbing states have attracted
a lot of attention by the theoretical physics community in recent
years~\cite{parity,hinrichsen}. For completeness we report here the
main results~\cite{nature}.

It is possible to show 
that on the symmetry line one can reformulate the problem in terms of
the \emph{dynamics of kinks} between patches of empty and occupied 
sites. 
Since the kinks are created and annihilated in pairs, the dynamics
conserves the initial number of kinks modulo two. In this way we can
present an exact mapping between a model with symmetric absorbing
states and one with parity conservation.

Outside the symmetry 
line the system belongs to 
the directed percolation universality class~\cite{percolation}.
We performed simulations starting either from
one and two kinks. In both cases $p_t=0.460(2)$, but the exponents 
were found to be different. Due to the conservation of the number
of kinks modulo two, starting from a single site one cannot observe
the relaxation to the absorbing state, and thus $\delta=0$. In this
case $\eta =0.292(5)$, $z=1.153(5)$. On the other hand, starting
with two neighboring kinks, we find $\eta=0.00(2)$,
$\delta=0.285(5)$, and $z=1.18(2)$. These results are consistent with
the parity conservation universality class~[3,4].

Let us now turn to the sensitivity of the model to a variation in the
initial configuration, i.e.\ to the study of \emph{damage spreading} or,
equivalently, to the location of the chaotic phase.
Given two replicas ${\boldsymbol x}$ and 
${\boldsymbol y}$, we define the difference 
${\boldsymbol w}$ as
${\boldsymbol w}={\boldsymbol x}\XOR{\boldsymbol y}$, where the symbol
$\XOR$ denotes the sum modulus two.
 
The damage $h$ is defined as the fraction of sites in
which $w=1$, i.e.\ as the Hamming distance between the configurations
${\boldsymbol x}$ and ${\boldsymbol y}$.
We study the case of maximal correlations by
using just one random number per site, corresponding to the smallest possible
chaotic region~\cite{damage}. 

In Fig.~3  the  
region in which the damage spreads is shown near the lower-right corner (chaotic domain). Outside this region
small spots appear near the phase boundaries,
due to the divergence of the relaxation time (second-order
transitions) or because a small difference in the initial
configuration can bring the system  to a different absorbing state
(first-order transition). The chaotic domain
is stable regardless of the initial density. On
the line $p_2=0$ the critical point of 
the density and that of the damage spreading
coincide.

\begin{figure}[t]
\centerline{\includegraphics[width=6cm]{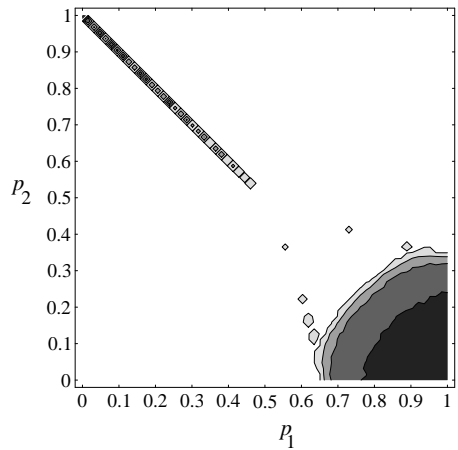}}
\caption{Phase diagram for the damage
found numerically by considering the evolution starting from
uncorrelated configurations with initial density equal to 0.5.}
\end{figure}
 
\subsection{Reconstruction of the potential}

An important point in the study of systems exhibiting absorbing states is 
the
formulation of a coarse-grained description using a \emph{Langevin
equation}. 
It is generally accepted that the directed percolation universal behavior is represented by 
\[
 \dfrac{\de c(x,t)}{\de t} =  
 a c(x,t) -b c^2(x,t) + \nabla^2 c(x,t) +
 \sqrt{c(x,t)}\alpha(x,t),
\]
where $c(x,t)$ is the density field, $a$ and $b$ are control parameters
and $\alpha$ is a Gaussian noise with correlations
$\langle \alpha(x,t) \alpha(x',t')\rangle =
\delta_{x,x'}\delta_{t,t'}$. The diffusion coefficient has been
absorbed into the parameters $a$ and $b$ and the time scale. 

It is possible to introduce a zero-dimensional approximation to the model by
averaging over the time and the space, assuming
that the system has entered a metastable state.  In this
approximation, the size of the original systems enters through the
renormalized coefficients  $\overline{a}$, $\overline{b}$,  
\[
 \dfrac{\de c(x,t)}{\de t} =  
 \overline{a} c(x,t) -\overline{b} c^2(x,t) +
 \sqrt{c(x,t)}\alpha(x,t),
\]
where also the time scale has been renormalized.

The associated Fokker-Planck equation is
\[
 \dfrac{\de P(c,t)}{\de t} = - \dfrac{\de}{\de c}(\overline{a}c
 -\overline{b}c^2)P(c,t)
 + \dfrac{1}{2} \dfrac{\de^2}{\de c^2}c P(c,t),
\]
where $P(c,t)$ is the probability of observing a density $c$ at
time $t$. One possible solution is a $\delta$-peak centered at the origin, 
corresponding to the absorbing state. 

By considering only those trajectories
that do not enter the absorbing state during the observation time, one
can impose a detailed balance condition, whose effective agreement
with the actual probability distribution has to be checked \textit{a
posteriori}.
The \emph{effective potential} $V$ is defined
as $V(c) = -\log(P(c))$ and can be found from
 the actual simulations.

\begin{figure}[t]
\begin{tabular}{cc}
\includegraphics[width=6cm]{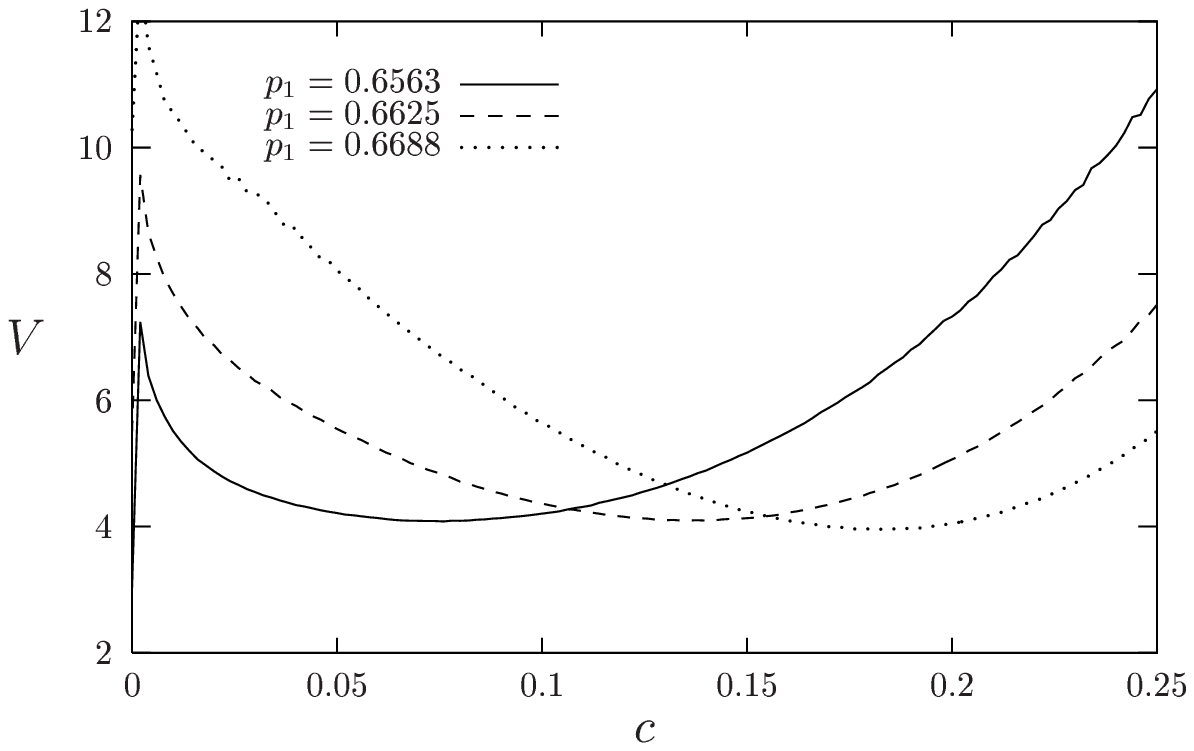} & 
\includegraphics[width=6cm]{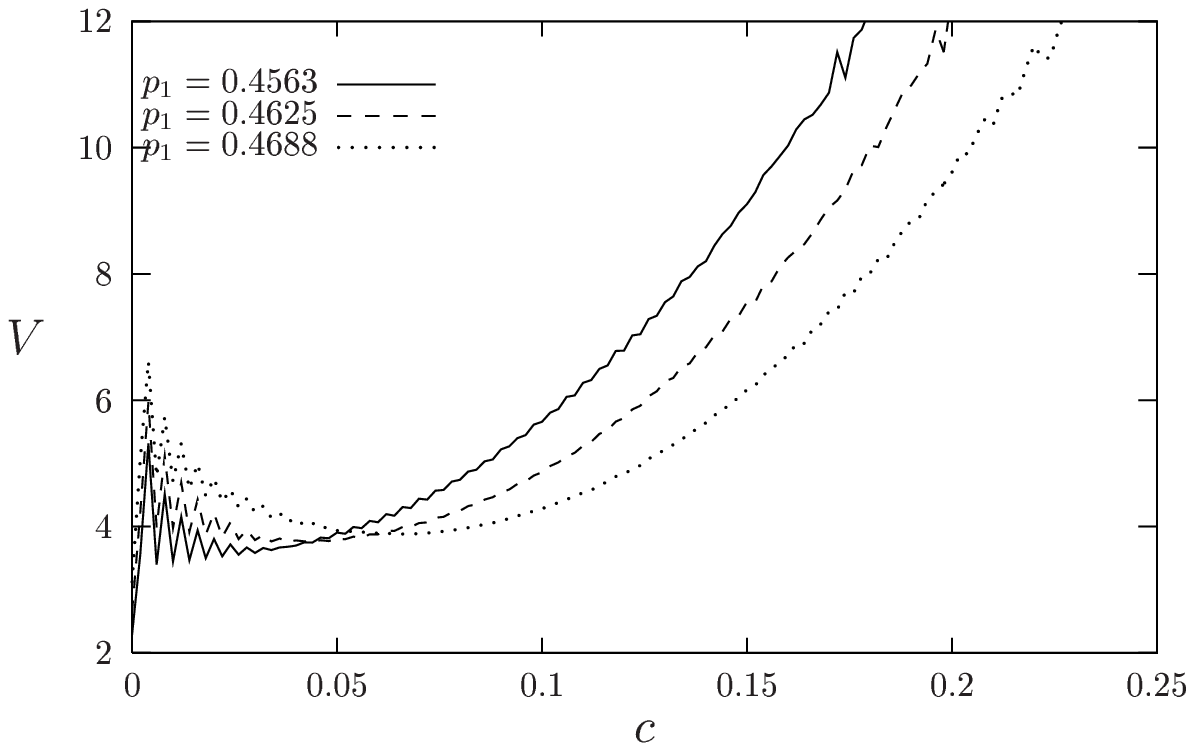}
\end{tabular}
\caption{Reconstruction of potential $V(c)$ 
for $p_2=0$ (left) and 
for the kink dynamics 
 on the line $p_2=1-p_1$ (right).}
\end{figure}

\begin{figure}[t]
\begin{tabular}{cc}
\includegraphics[width=6cm]{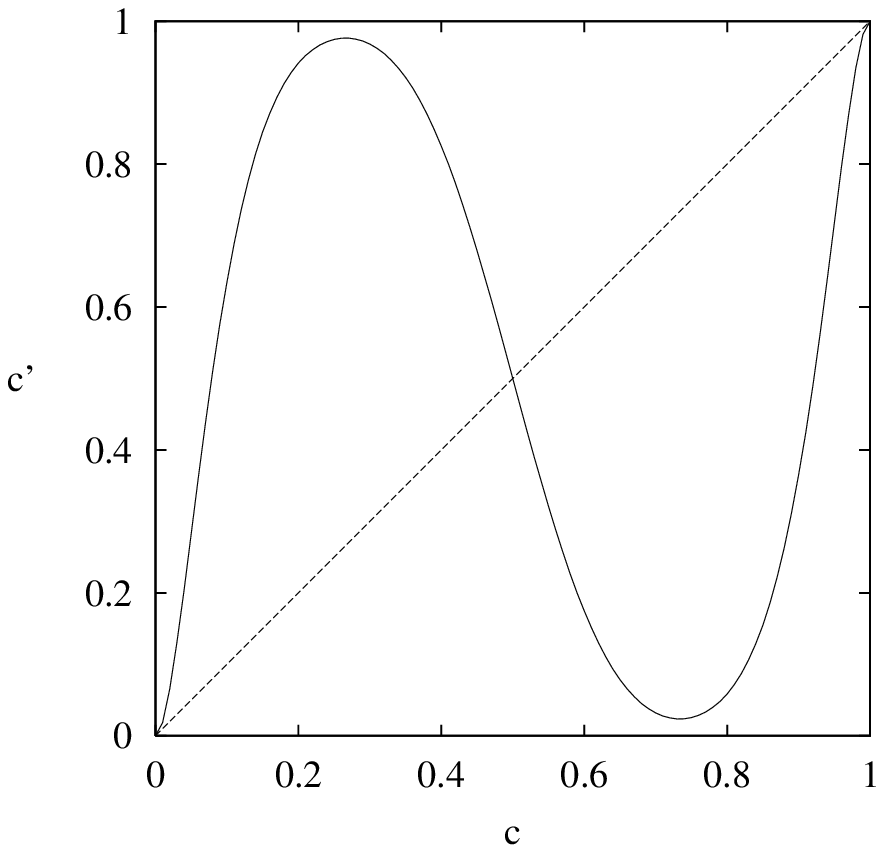} & 
\includegraphics[width=6cm]{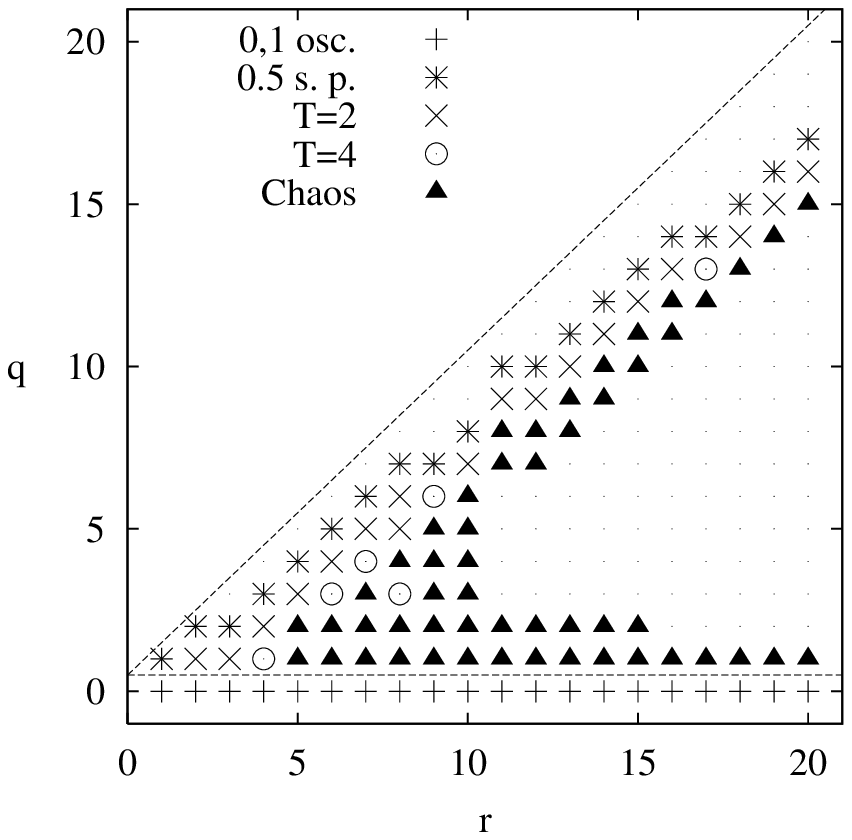}
\end{tabular}
\caption{\label{fasecm} Case $H=0$, $J=-\infty$: the 
mean field map
Eq.~\protect\eqref{cm} (left) for
$r=10$ and $q=2$ and the
mean-field $r-q$ phase diagram (right). In the phase diagram the absorbing
states are always present. Points mark parameter values for which the
absorbing states are  the only stable attractors. 
A plus sign denotes period-2 temporal 
oscillations between absorbing states, 
a star denotes the presence of a stable
point at $c=0.5$, a cross (circle) denotes period-two (four)
oscillations between two non-zero and non-one densities, triangles
denote chaotic oscillations.} 
\end{figure}

In the left panel of Fig.~4 we show the profile of the reconstructed
potential $V$ for some values of $p$ around 
the critical value
on the line $p_2=0$, 
over which the model belongs to the DP
universality class. One can observe that the curve becomes broader in
the vicinity of the critical point, in correspondence of the
divergence of critical fluctuations $\chi \sim |p-p_c|^{-\gamma'}$,
$\gamma'=0.54$~\cite{reconstruction}. By repeating the same type of
simulations for the kink dynamics (random initial condition), we
obtain slightly different curves, as shown in the right panel of Fig.~4. We
notice that all curves have roughly the same width. Indeed, the
exponent $\gamma'$ for systems in the PC universality class is
believed to be exactly $0$~\cite{jensen}, as given by the scaling
relation $\gamma' = d \nu_\perp - 2\beta$~\cite{reconstruction}.
Clearly, much more information
can be obtained from the knowledge
of $P(c)$, either by direct numerical simulations or dynamical mean
field trough finite scale analysis, as shown for instance in
Ref.~\cite{Jensen1}.

\section{Larger neighborhoods}
\label{sec:meanfield}

In order to study the effects of a larger neighborhood and
different threshold values $q$, 
let us start with the well known two-dimensional ``Vote''
model. It is defined on a square lattice, with a Moore
neighborhood composed by 9 neighbors, synthetically denoted M in the following~\cite{vichniac}. If a strict majority rule $q=4$
is applied (rule M56789, same convention as in Sec.~\ref{sec:model}) to a random initial
configuration, one observes the rapid
quenching of small clusters of ones and zero, similar to what happens
with rule
T23 in the one-dimensional case. 
A small noise quickly leads the system to an absorbing state. 
On the other hand, a small frustration $q=3$ (rule M46789) for an
initial density $c^0=0.5$ leads to a metastable point formed by
patches that evolve very slowly towards one of the two absorbing
states. However, this metastable state is given by the perfect balance
among the absorbing states. If one starts with a different initial
``magnetization'', one of the two absorbing phases quickly dominates, except for
small imperfections that disappear when a small noise is added.

In the general case, the mean-field equation is
\begin{equation}\label{cm}
	c' = \sum_{s=0}^{2r+1} \binom{2r+1}{s} c^s (1-c)^{2r+1-s} p_s,
\end{equation}
sketched in the left panel of Fig.~\ref{fasecm}.

We studied the asymptotic behavior of this map 
for different values of $r$ and
$q$. For a given $r$ there is always a critical $q_c$ value of $q$ for
which the active phase disappears, with an approximate correspondence
$q_c \simeq 4/5 r$. The active phase 
 is favored by the presence of frustrations and
the absence of the external pressure,
i.e.\ for $J<0$ and $H=0$. We performed extensive computations for the
extreme case $J=- \infty $, $H=0$ corresponding to a society of
non-conformists without television. As shown in the right panel of 
Fig.~\ref{fasecm},
by increasing the neighborhood size $r$, one starts observing oscillations
in the equilibrium process. This is
evident in the limit of infinite neighborhood: the parallel dynamics
induced by elections (in our model) makes individual tend to disalign
from the marginal majority, originating temporal oscillations that can
have a chaotic character. Since the absorbing states are present, and
they are stable for $q>0$, the coherent oscillations of the population
can bring the system into one absorbing state. This is reflected by
the ``hole'' in the active phase in the mean field phase diagram. 

Preliminary lattice simulations (not shown) reveal that this behavior is still present if there is a sufficiently large fraction of long-range connections due to the small-world effect~\cite{small},
while the active phase is compact in the short-range case.

\section{Conclusions}
Although this model is quite rough,
there are aspects that present some analogies with the behavior 
of a real society. In particular, the role of education, represented
by the $J$ parameter. In a society dominated by conformists,
the transitions are difficult and catastrophic, while in the opposite case
of non-conformist people the transitions are smooth. However, in the
latter case a great social pressure is needed to gain the majority. 

On the other hand, if the neighborhood is large and non-conformism is
present, one can observe the phenomenon of political instabilities
given by temporal oscillations of population opinion, which could be
interpreted as a symptom of healthy democracy.

\end{document}